\documentclass[prl,twocolumn,showpacs,superscriptaddress,floatfix]{revtex4}
\usepackage{graphicx}
\usepackage{dcolumn,float}
\usepackage{amsmath,amssymb}
\usepackage{subfigure}
\usepackage{hyperref}

\newcommand{\be}{\begin{eqnarray}}

\newcommand{\ee}{\end{eqnarray}}

\newcommand{\qq}{\begin{eqnarray}}
\newcommand{\qqq}{\end{eqnarray}}

\begin{document}

\title{Universal spatial correlations in random spinor fields}

\author{Juan Diego Urbina}
\affiliation{Institut f\"ur Theoretische Physik, Universit\"at Regensburg, 93040 Regensburg, Germany}
\author{Michael Wimmer}
\affiliation{Instituut-Lorentz, Universiteit Leiden, P.~O.~Box 9506, 2300 RA Leiden, The Netherlands}
\author{Dominik Bauernfeind}
\affiliation{Institut f\"ur Theoretische Physik, Universit\"at Regensburg, 93040 Regensburg, Germany}
\author{Diego Espitia}
\affiliation{Universidad Pedagogica y Tecnologica de Colombia, UPTC, Tunja, Colombia}
\author{Inanc Adagideli}
\affiliation{Faculty of Engineering and Natural Sciences, Sabanci University, Orhanli-Tuzla, 34956 Istambul, Turkey}
\author{Klaus Richter}
\affiliation{Institut f\"ur Theoretische Physik, Universit\"at Regensburg, 93040 Regensburg, Germany}

\begin{abstract}
We identify universal spatial fluctuations in systems with non trivial spin dynamics. To this end we calculate by exact numerical diagonalization a variety of experimentally relevant correlations between spinor amplitudes, spin polarizations and spin currents both in the bulk and near the boundary of a confined two-dimensional clean electron gas in the presence of spin-orbit interaction and a single magnetic impurity. We support or claim of universality with the excellent agreement between the numerical results and system-independent spatial correlations of a random field defined on both the spatial and spin degrees of freedom. A rigorous identity relating our universal predictions with response functions provides a direct physical interpretation of our results in the framework of linear response theory.  
\end{abstract}

%\pacs{71.10.Hf, 75.10.Jm, 75.30.Gw, 74.20.Mn}
% PACS, the Physics and Astronomy Classification Scheme.

\maketitle

\newcommand{\bb}{\boldsymbol{\beta}}
\newcommand{\ba}{\boldsymbol{\alpha}}

%%%%%%%%%%%%%%%%1%%%%%%%%%%%%%%%%%%%%%%%%%%%%%%%%%%%%%%%%%%%%%%
The field of spintronics, which deals with the use of the electron spin of freedom as means to transmit, store and process energy and information, has experienced impressive advances during the last decade \cite{in1}. The possibility of manipulating spin densities and currents by means of their correlations with the easier to control charge (and electrical current) degrees of freedom has been in the focus of semiconductor-based spintronics. Here the spin Hall effect \cite{in2}, the creation of a spin imbalance across a sample generated by a charge current in the presence of spin-orbit interaction (SOI), is a primary example of such spatial correlation and the close connection between the spin and charge degrees of freedom. 

In this context, universality of spatial correlations can arise from two basic mechanisms. In systems where the mean free path $l_{f}$ is much smaller than the system size $L$, average over the random distribution of obstacles produces results which are largely independent of the geometry of the confinement. This diffusive limit has been extensively studied using diagramatic techniques based on disorder average \cite{dis} which, however, cannot deal with ballistic systems where $l_{f}/L \gg 1$, a regime which is now easily achieved in high-mobility semiconductor 2D electron gases \cite{in3}. In this ballistic case universality arises due to electron scattering with the irregular boundaries, namely, from the presence of classical chaos. It is in this regime where semiclassical approaches to universality \cite{sem1} (and its breakdown \cite{sem2}) in spin and charge transport in the presence of SOI have been very succesful. 

Spatial fluctuations in ballistic spinor  systems have been addressed in \cite{malo} in the limit of vanishing SOI, and recently in \cite{ulloa} the spatial correlations of charge densities for the bulk have been studied by means of Random Matrix Theory. The extension to systems with non-zero local spin polarization (which is realized by an STM tip, for example) and in the presence of boundaries requires substantial technical and conceptual steps beyond Refs. \cite{malo,ulloa}. Our goal is to fill this gap. 

Our starting point is the Schr\"odinger equation ($\hat{I}(\hat{1})$ is the unit operator in spin (position) space) 
\begin{equation}
\label{eq:Schr0}
\left[\frac{\hat{{\bf p}}^ {2}}{2m} \otimes {\hat I}+{\hat 1}\otimes\frac{\hbar k_{{\rm s.o}}}{m}\left( \hat{\sigma}_{x}\hat{p}_{y}-\hat{\sigma}_{y}\hat{p}_{x}\right)\right]|\Psi_{n} \rangle=E_{n}|\Psi_{n} \rangle,
\end{equation}
with boundary conditions 
\begin{equation}
\label{eq:Schr}
\Psi_{n}({\bf r} \in \partial \Omega,s)=0 {\rm \ and \ }\Psi_{n}({\bf q},s)=\delta_{s,\uparrow},
\end{equation}
where we use $\Psi_{n}({\bf r},s)=(\langle {\bf r}|\otimes \langle s|)|\Psi_{n} \rangle$ and $\delta_{s,s'}$ is the kronecker delta. Eqs. (\ref{eq:Schr0},\ref{eq:Schr}) describe the dynamics of an electron with effective mass $m$ inside a quantum dot with Dirichlet boundary conditions along $\partial \Omega$ in the presence of SOI with inverse precession length $k_{{\rm s.o}}$, and of a magnetic impurity (polarized in the $|s=\uparrow \rangle$ direction) located at position ${\bf q}$. The momentum operator $\hat{{\bf p}}=-i\hbar(\partial/ \partial x,\partial/ \partial y)$ acts on the orbital degrees of freedom ${\bf r}=(x,y)$, and $(\hat{\sigma}_{x},\hat{\sigma}_{y},\hat{\sigma}_{z})$ are the Pauli matrices acting on the space spaned by the eigenstates of $\hat{\sigma}_{z}$ (denoted by $|s\rangle$ with $s=\uparrow, \downarrow$).  

\begin{figure}[ht]
\includegraphics[width=0.7\columnwidth,height=0.4\columnwidth,clip,angle=0]{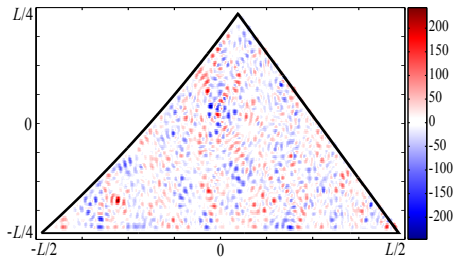}
\caption{Spin polarization in $z$-direction in units of $L^{-2}$ for the linear combination of the two degenerate states $n=400$ of the "star billiard" such that the combined state is polarized in the $|\uparrow \rangle$ direction at the origin ($k_{\rm s.o}L=10$).}     
\label{fig:fig1}
\end{figure}

Time Reversal Invariance (TRI) is expressed by the condition \cite{sak} $\hat{H}\hat{T}=\hat{T}\hat{H}$, where $\hat{T}= -i \hat{\sigma}_{y} \hat{K}$ is the time reversal operator and $\hat{K}$ indicates complex conjugation in the eigenbasis of $\hat{\sigma}_{z}$. Since Dirichlet boundary conditions are also invariant under time reversal, without a magnetic impurity eigenstates of $\hat{H}$ come in degenerate (Kramers) pairs $(|\Psi_{n} \rangle,|\Psi_{n}^{T} \rangle=\hat{T}|\Psi_{n} \rangle)$. Kramers degeneracy prevents the very existence of non-zero local spin polarization if the state of the system is given by an incoherent (statistical) superposition,  
\begin{equation}
{\rm Tr ~} \hat{\rho}_{n} \hat{\sigma}_{j}=0 {\rm \ \ for \ \ }\hat{\rho}_{n}\sim |\Psi_{n} \rangle \langle\Psi_{n}|+|\Psi_{n}^{T} \rangle \langle\Psi_{n}^{T}|.
\end{equation}
This is the reason why Ref.~\cite{ulloa} deals only with correlations of the local charge, as this is the only density that does not vanish when averaged over the Kramers pair. However, any specific coherent superposition $a|\Psi_{n}\rangle+b|\Psi_{n}^{T}\rangle$ shows indeed a very rich spatial and spin structure, which is washed out if we neglect coherences.  Physical realizations of such coherent superpositions are achieved by breaking TRI, and in particular by introducing spin selective boundary conditions, as in Eq. (\ref{eq:Schr}). As an example, Fig.~(\ref{fig:fig1}) shows the spin polarization in $z$-direction for a typical eigenstate of our hamiltonian where the magnetic impurity is localized inside of a ballistic cavity. Here we will present a theoretical approach to understand and predict the spatial statistics of such imprinted pattern.

\begin{figure}[htbp]
\includegraphics[width=0.48\textwidth]{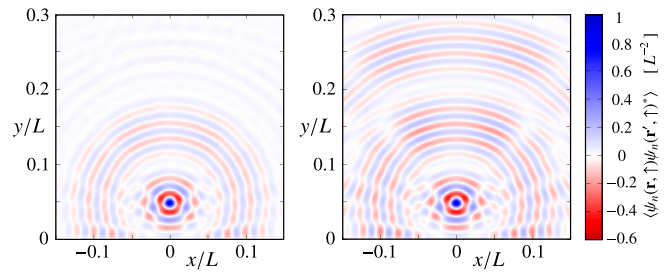} 
\caption{Comparison between the exact (left panel) spatial two-point correlation function $\langle \psi_{n}(\mathbf r,\uparrow) \psi_{n}(\mathbf r',\uparrow)^{*}\rangle$ (defined by impurity and energy average over 300 states near $n=3000$) as a function of $\mathbf{r}$, near a wall at $x=0$ (with $\mathbf r'=(0,0.05L)$) , and the universal results $C^{\uparrow,\uparrow}_{0}(\mathbf r, \mathbf r';E) + C^{\uparrow,\uparrow}_{1}(\mathbf r, \mathbf r';E)$ (right), see text. We use $k_{\rm s.o}/k=0.1$ and $kL=300$.}     
\label{fig:fig2}
\end{figure}

For the numerical simulations we consider the desymmetrized "star billiard" (shown in Fig.~\ref{fig:fig1}), which is characterized by only one parameter (the radius of the circular arc in units of the length $L$ of the horizontal segment). For $k_{\rm s.o}=0$ this billiard is known to display hard chaos in the classical limit and therefore it is expected to exhibit universality in the spatial correlations of its quantum eigenstates \cite{berr}. In order to construct local observables, we explicitely diagonalize the hamiltonian using the method reported in \cite{mike}, which gives the corresponding Kramers pair for each eigenenergy $E_{n}$ . For a given position ${\bf q}$ of the spin impurity the (now unique) eigenstate of the system is given by the linear combination
\begin{equation}
|\psi_{n}^{a,b}\rangle=a({\bf q})|\Psi_{n}\rangle+b({\bf q})|\Psi_{n}^{T}\rangle,
\end{equation}
where the coefficients $a({\bf q}), b({\bf q})$ must fulfil $|a|^{2}+|b|^{2}=1$ and are adjusted such that $\Psi_{n}({\bf q},s)=\delta_{s,\uparrow}$. For fixed $n$ any local function $ {\cal F }(\psi_{n}^{a,b}({\bf r},s),\psi_{n}^{a,b}({\bf r}',s'))$  of the state will fluctuate when ${\bf q}$ is randomly choosen inside the billiard. This quasi-random character of the spatial fluctuations is used to replace the average over the impurity position ${\bf q}$ by an average over $(a,b)$ on the unit sphere,
\begin{eqnarray}
\int_{\Omega}&&{\cal F }(\psi_{n}^{a({\bf q}),b({\bf q})}({\bf r},s),\psi_{n}^{a({\bf q}),b({\bf q})}({\bf r}',s'))d{\bf q}\\
&&=\int_{|a|^{2}+|b|^{2}=1}{\cal F }(\psi_{n}^{a,b}({\bf r},s),\psi_{n}^{a,b}({\bf r}',s'))da db. \nonumber 
\end{eqnarray}
This is the way we use the numerical Kramers pair $(|\Psi_{n} \rangle,|\Psi_{n}^{T} \rangle)$ to construct the expectation values of local observables ${\cal F }$ at fixed energy. Now we proceed in the usual way one studies spatial fluctuations of wavefunctions in classically chaotic systems. We use the exact numerical results for local observables to perform an energy average where we expect universality to emerge. That this is actually the case can be seen in Fig.~\ref{fig:fig2}, where the two-point spatial correlation  $\langle \psi_{n}^{a,b}({\bf r},s)\psi_{n}^{a,b}({\bf r}',s')^{*}\rangle$  of the spinor amplitude (calculated by impurity and energy average of the exact eigenfunctions) is compared with the universal prediction of the Spinor Random Wave Model (SRWM) to be presented bellow.

Any theory which attempts to describe the universality of spatial correlations must deal with two different aspects: first, the issue of the universal behaviour of the amplitude correlator both in the bulk and  near a boundary, and second the appropiate description of correlators beyond the bilineal form in the amplitudes. The original approaches to universal spatial fluctuations in chaotic systems considered these two phenomena to share the same origin (as both can be derived from Berry's ansatz stating that wavefunctions of classically chaotic systems behave as random superpositions of plane waves \cite{berr}). Later it was recognized \cite{corr} that the two-point correlator is actually an intrinsically microscopic object that can be derived without any further assumption from the exact Green function $\hat{G}^{\pm}(E)=(\hat{H}-E {\pm}i 0^{+})^{-1}$ by means of the formula
\begin{equation}
\label{eq:univ}
 C(\mathbf r, \mathbf r';E) \simeq\frac{1}{2\pi i}\frac{1}{\varrho(E)}\langle G^-(\mathbf r', \mathbf r;E)-  G^+(\mathbf r, \mathbf r';E) \rangle.
\end{equation}
Here $\langle \ldots \rangle$ denotes impurity and energy average (over a small window around $E$) and $\varrho(E)$ is the smooth part of the density of states. For systems with SOI and away from the boundary, the exact Green function is approximated by its bulk value to get the universal prediction for the bulk,
\begin{equation}
\label{eq:C0}
 C_0(\mathbf r, \mathbf r';E)=\frac{1}{4 k} \left(
    \begin{array}{cc}
            C_D(d;E) &  -e^{-i\theta} C_S(d;E)  \\
            e^{i\theta} C_S(d;E) &  C_D(d;E)
          \end{array}
        \right),
\end{equation}
where $C_D(d;E) = k_+J_0(k_+ d)+k_-J_0(k_- d)$, $C_S(d;E) = k_+J_1(k_+ d)-k_-J_1(k_- d)$ and $J_{n}(x)$ are Bessel functions. We further defined $k_{\pm}=\sqrt{k^{2}+k_{\rm s.o}^{2}}\pm k_{\rm s.o}$ with $k=\sqrt{2mE/ \hbar^{2}}$ and $\mathbf r-\mathbf r'=d(\cos{\theta},\sin{\theta})$. The result (\ref{eq:C0}) can be obtained using a modified Berry ansatz including SOI, as in Refs.~\cite{ulloa,chris}. 

To go beyond the results for the bulk we use a multiple reflection expansion to construct the matrix-valued Green function near a wall, assumed to be an infinite straight line at $x=0$. Translational invariance in the $x$-direction suggests to perform a Fourier transform (indicated by a tilde) from $x-x'$ to $k_{x}$. The method provides a closed form for the boundary contribution as (for notational convenience the dependence with $E$ is kept implicit)
\begin{eqnarray}
\label{Gamma}
\tilde {G_1}(k_x,y,y')=2{\partial \tilde G_0}(k_x,y)(1-2{\partial \tilde G_0}(k_x,0))^{-1} {\tilde G_0}(k_x,y'), \nonumber
\end{eqnarray}
in terms of the bulk Green function 
\begin{eqnarray}
\label{mixedGreen}
{ \tilde G_0}(k_x,y-y') = \sum_{+,-}(\pm)\frac{e^{-a_{\pm}|y-y'|}}{2a_{\pm}}\left(f_{\pm}\hat{I}-g\hat{\sigma}_y+h \hat{\sigma}_x\right), \nonumber
\end{eqnarray}
and its normal derivatives ${\partial \tilde G_0}$ at the boundary. We introduced $f_{\pm}=g(k^2-k^{2}_{\pm})/k_{x}$, and $h=iga_{\pm} \mbox{sgn}(y-y')$ with $g= k_x/(2\sqrt{k^2+k_{so}^2})$ and $a_{\pm}=\sqrt{k_{x}^{2}-k_{\pm}^{2}}$ ($\Re a_{\pm} >0$). 

In order to construct the correlation function in real space, the inverse Fourier transform of $\tilde {G_1}$ is calculated in stationary phase approximation, well justified in the regime $kL \gg 1$. We obtain a linear combination of expressions with phases $\Phi(k_x) = \sqrt{k^2_\pm-k_x^2}\cdot y+\sqrt{k^2_\pm-k_x^2}\cdot y'+k_x(x-x')$,  showing that the geometry of the saddle points $\Phi'(k_x)=0$ is a deformed version of the Snell law, with two (instead of just one) possibilities for the incoming and outgoing wavevectors. The limit $k_{{\rm s.o}}/k \to 0,k_{{\rm s.o}}L \to {\rm const.}$ where the SOI is not considered for the stationary phase condition gives insufficient results for the spatial correlations, and we use instead a consistent solution for each independent combination of beams up to second order in $k_{{\rm s.o}}/k$. The effect on the spatial correlations is then incorporated as a contribution $C_{1}$ on top of the bulk result $C_{0}$, Eq.~(\ref{eq:C0}). The effect of a nearby boundary on the two-point amplitude correlator is depicted in Fig.~\ref{fig:fig2} showing excellent agreement between numerical simulations and our analytical (but very lengthy) formula. This boundary effect is the spatial analogue of the perimeter term in the Weyl formula \cite{sto} for the smooth part of the density of states of a system with SOI \cite{dom}. 

We have checked that changes on the size of the energy window, relative positions $\mathbf r, \mathbf r '$, position of the wall, SOI strength do not affect the quality of our results for all the entries of the correlation matrix. Therefore we conclude that spatial correlations of spinor amplitudes are described by the formula (\ref{eq:univ}), and universality emerges when the Green function can be approximated by its universal limit for the bulk or near a hard wall. That this is precisely the case for chaotic quantum systems was shown for the scalar case in \cite{urb2}, and the same argument (that paths with multiple reflections produce sub-dominant effects) holds here. 

We now turn our atention to the experimentally more relevant case of spatial correlations for local densities. We consider observables of the form
\begin{equation}
\hat{A}^{d}({\bf r})=\delta(\hat{\bf r}-{\bf r}) \otimes \hat{A}.
\end{equation}
For given position ${\bf r}$, the choice $\hat{A}=\hat{1}$ describes the local charge density, while $\hat{A}=\hat{\sigma}_{i}$ gives the spin density in $i$-th direction. Introducing the spinor $\psi^{a,b}_{n}({\bf r})=\langle {\bf r}|\psi^{a,b}_{n}\rangle$, the numerical spatial density correlations are then constructed from the numerical eigenstates by impurity and energy average 
\begin{eqnarray}
\label{eq:ex}
&&C_{AB}({\bf r},{\bf r}';E)= \\ &&\langle\left[\psi^{a,b}_{n}({\bf r})^{\dagger}\hat{A}\psi^{a,b}_{n}({\bf r})\right]\left[\psi^{a,b}_{n}({\bf r}')^{\dagger}\hat{B}\psi^{a,b}_{n}({\bf r}')\right]\rangle. \nonumber
\end{eqnarray}

Contrary to the amplitude correlators, density correlations are not bilineal in the components of the state and therefore cannot be directly related with the Green function. Following a well established procedure in systems without spin, we assume that the spinor amplitudes have {\it Gaussian fluctuations} \cite{corr}, and we replace the energy and impurity averages by a single average over a functional distribution of spinor fields,
\begin{eqnarray}
\label{eq:exg}
C_{AB}^{SRWM}({\bf r},{\bf r}';E)=\int P(\psi)\psi({\bf r})^{\dagger}\hat{A}\psi({\bf r})\psi({\bf r}')^{\dagger}\hat{B}\psi({\bf r}') {\cal D}[\psi]. \nonumber
\end{eqnarray} 
The probability distribution $P(\psi)$ is Gaussian and therefore uniquely given by its two-point correlation function $\int P(\psi) \psi({\bf r}) \psi({\bf r})^{\dagger}{\cal D}[\psi]=C(\mathbf r, \mathbf r';E)$, which we replace by the universal amplitude correlator based on the microscopic Green function. 

\begin{figure}[htbp]
\includegraphics[width=0.48\textwidth]{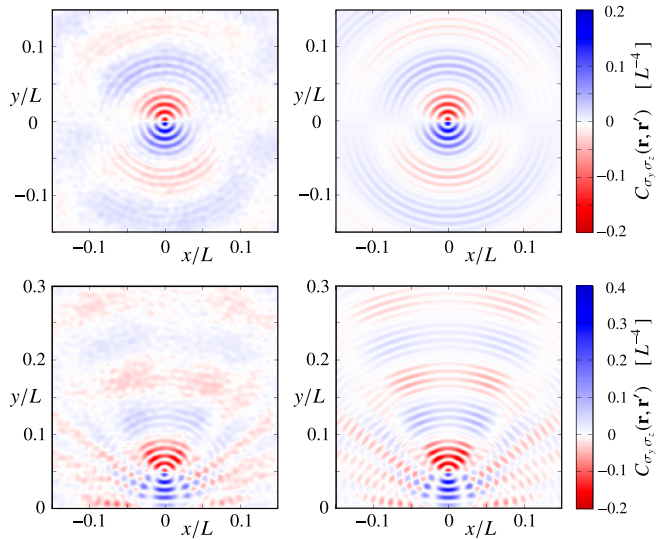}
\caption{Spatial correlation $C_{\sigma_{y} \sigma_{z}}(\mathbf r, \mathbf r';E)$ of spin densities in the bulk (top) and near a wall at $x=0$ (bottom) as function of ${\mathbf r}$. Left panels: results of Eq.~(\ref{eq:ex}) based on numerically obtained eigenstates. Right panels: universal prediction $C_{\sigma_{y} \sigma_{z}}^{SRWM}(\mathbf r, \mathbf r';E)$, Eq. (\ref{eq:RWM}). We use ${\mathbf r'}=(0,0.05L)$, $k_{\rm s.o}/k=0.1$ and $kL=300$. Averages are calculated using $300$ Kramers pairs around $E_{3000}$.}     
\label{fig:fig3}
\end{figure}

\begin{figure}[htbp]
\includegraphics[width=0.48\textwidth]{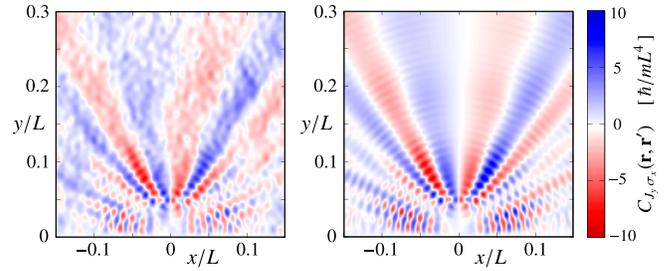} 
\caption{Spatial correlation $C_{J_{y} \sigma_{z}}(\mathbf r, \mathbf r';E)$ between spin current and spin density  near a wall at $x=0$ produced, for instance, by applying a spin current with $y$ polarization at ${\mathbf r'}=(0,005L)$ and measuring the dependence with ${\mathbf r}$ of the spin polarization in $z$ direction. Left side: numerical calculation from Eq.~(\ref{eq:ex}). Right side: universal prediction $C_{J_{y} \sigma_{z}}^{SRWM}(\mathbf r, \mathbf r';E)$ from Eq. (\ref{eq:RWM}). The parameters are the same as in Fig.~\ref{fig:fig3}.}     
\label{fig:fig4}
\end{figure}

Having at hand a Gaussian theory with known two-point correlators, we can decouple averages over higher order functionals of the state by straightforward use of Wick's theorem. For the particular case of local observables, this easily gives \cite{dom}
\begin{eqnarray}
\label{eq:RWM}
&&C_{AB}^{SRWM}({\bf r},{\bf r}';E)= {\rm Tr}\left[\hat{A} C(\mathbf r, \mathbf r';E)\hat{B}C(\mathbf r', \mathbf r;E)\right] \nonumber \\
&&+{\rm Tr}\left[\hat{A} C(\mathbf r, \mathbf r';E)\right]{\rm Tr}\left[\hat{B}C(\mathbf r', \mathbf r;E)\right] 
\end{eqnarray}
where the trace is over spin variables only. Eq.~(\ref{eq:RWM}) is the main result of this paper. It allows us to derive universal results for spatial correlations of local observables after inserting the expressions $C_{0}, C_{1}$ of the correlator $C(\mathbf r, \mathbf r';E)$ for the bulk or its modification near the wall. In order to check the underlying Gaussian assumption, in Figs.~\ref{fig:fig3},\ref{fig:fig4} we compare the result (\ref{eq:RWM}) with numerical results based on Eq. (\ref{eq:ex}). We find considerable agreement even for the subtle patterns emerging from interference effects due to the boundary. 

The physical relevance of $C_{AB}$ can be extended beyond its statistical interpretation by noticing that the connected part of $C_{AB}^{SRWM}$ at the Fermi energy $E=E_{F}$ can be rigorously related through
\begin{eqnarray}
\label{eq:LR}
\lim_{w \to 0}&&\int_{-\infty}^{\infty}\phi_{AB}(t)\frac{{\rm e~}^{i w t}}{w}dt \\ &&=i{\rm Tr}\left[\hat{A} C(\mathbf r, \mathbf r';E_{F})\hat{B}C(\mathbf r', \mathbf r;E_{F})\right] \nonumber 
\end{eqnarray}  
to the d.c.~component of the dynamical response function $\phi_{AB}(t-t')\propto \langle [\hat{A}^{d}({\bf r},t),\hat{B}^{d}({\bf r},t')] \rangle$ describing the change of the expectation value of $\hat{A}^{d}$ at time $t$ when an infinitesimal perturbation affects the system at time $t'$ through a coupling with the observable $\hat{B}^{d}$ \cite{Kubo}. Eq.~(\ref{eq:LR}) relates an experimentally accessible quantity, the response function, with the correlator quantifying the statistical fluctuations of the random spinor field.

To summarize, we have shown that electrons in confined chaotic geometries in the presence of spin-orbit coupling exhibit spatial spin and spin current correlations given by the universal correlations of a Gaussian random spinor field. Our results can be applied to a large class of correlators both in the bulk and near a boundary. They hold not only for the correlations between amplitudes but also for spatial correlations of spin densities and spin currents, more generally for any pair of local observables. A rigorous identity relating linear response coefficients to these universal correlators gives further insight into our results and opens a straightforward possibility to measure such correlations.   

{\it Acknowledgements}. This work was supported by the Deutsche Forschungsgemeinschaft within SFB 689 (JDU, KR).

%%%%%%%%%%%%%%%%%%%%%%%%%%%%%%%%%%%%%%%%%%%%%%%%%%%%%%%%%%%%%%

%%%%%%%%%%%%%%%%%%%%%%%%%%%%%%%%%%%%%%%%%%%%%%%%%%%%%%%%%%%%%%

\end{document}